\documentclass[pra,onecolumn,floatfix,superscriptaddress,longbibliography,notitlepage, nofootinbib]{revtex4-1}
\pdfoutput=1

\usepackage[utf8]{inputenc}
\usepackage{braket}
\usepackage{amsmath}
\DeclareMathOperator{\Tr}{Tr}
\usepackage{dcolumn}   
\usepackage{bm}        
\usepackage{amssymb}
\usepackage{mathtools}
\usepackage{tikz}
\usepackage{qcircuit}
\usepackage{appendix}
\usepackage{hyperref}
\usepackage{subcaption}
\usepackage{amsmath,amsfonts,amssymb}
\usepackage{mathtools}
\usepackage{thmtools,thm-restate}
\usepackage{algorithm}
\usetikzlibrary{3d}
\usetikzlibrary{decorations.markings}

\usepackage[T1]{fontenc}
\usepackage{pst-3dplot, pst-math}

\usepackage{algpseudocode}

\newtheorem{definition}{Definition}
\newtheorem{theorem}{Theorem}
\newtheorem{lemma}{Lemma}

\usetikzlibrary{arrows.meta}

\def\BibTeX{{\rm B\kern-.05em{\sc i\kern-.025em b}\kern-.08em
    T\kern-.1667em\lower.7ex\hbox{E}\kern-.125emX}}

\begin{document}
\title{Quantum Algorithm For Solving Nonlinear Algebraic Equations }
\author{Nhat A. Nghiem}
\affiliation{Department of Physics and Astronomy, State University of New York at Stony Brook, Stony Brook, NY 11794-3800, USA}
\affiliation{C. N. Yang Institute for Theoretical Physics, State University of New York at Stony Brook, Stony Brook, NY 11794-3840, USA}
\author{Tzu-Chieh Wei}
\affiliation{Department of Physics and Astronomy, State University of New York at Stony Brook, Stony Brook, NY 11794-3800, USA}
\affiliation{C. N. Yang Institute for Theoretical Physics, State University of New York at Stony Brook, Stony Brook, NY 11794-3840, USA}
\affiliation{Institute for Advanced Computational Science, State University of New York at Stony Brook, Stony Brook, New York 11794-5250, USA}
\begin{abstract}
Nonlinear equations are challenging to solve due to their inherently nonlinear nature. As analytical solutions typically do not exist, numerical methods have been developed to tackle their solutions. In this article, we give a quantum algorithm for solving a system of nonlinear algebraic equations, in which each equation is a multivariate polynomial of known coefficients. Building upon the classical Newton method and some recent works on quantum algorithm plus block encoding from the quantum singular value transformation, we show how to invert the Jacobian matrix to execute Newton's iterative method for solving nonlinear equations, where each contributing equation is a homogeneous polynomial of an even degree. A detailed analysis are then carried out to reveal that our method achieves polylogarithmic time in relative to the number of variables. Furthermore, the number of required qubits is logarithmic in the number of variables. In particular, we also show that our method can be modified with little effort to deal with polynomial of various types, thus implying the generality of our approach. Some examples coming from physics and algebraic geometry, such as Gross-Pitaevski equation, Lotka-Volterra equations, and intersection of algebraic varieties, involving nonlinear partial differential equations are provided to motivate the potential application, with a description on how to extend our algorithm with even less effort in such a scenario. Our work thus marks a further important step towards quantum advantage in nonlinear science, enabled by the framework of quantum singular value transformation. 
\end{abstract}
\maketitle

\section{Introduction}
Quantum mechanics has offered a new computational paradigm, i.e., quantum computation~\cite{feynman2018simulating, deutsch1985quantum}. Following some of earlier landmark contributions, such as probing properties of black-box functions~\cite{deutsch1992rapid}, Shor's factoring algorithm~\cite{shor1999polynomial}, and Grover's search algorithm~\cite{grover1996fast},  tremendous efforts have been made to uncover further quantum advantage. One of the exciting results along the line is the quantum algorithm for solving linear systems of equations, or the HHL algorithm~\cite{harrow2009quantum}. Their work suggests a new path for achieving quantum advantage, in particular, showing that matrix inversion is BQP-complete. Additionally, the techniques used in~\cite{harrow2009quantum} have been modified in~\cite{wiebe2012quantum} to perform data fitting, which is also a very challenging computational problem. More recently, those techniques have been employed to solve gradient descent problems~\cite{rebentrost2019quantum}, which yields exponential speedup with respect to the input size. Aside from~\cite{harrow2009quantum}, another seminal work includes the quantum algorithm for principle component analysis~\cite{lloyd2013quantum}. A very important routine proposed in~\cite{lloyd2013quantum} is the exponentiation of a density operator (non-negative operator with unit trace). As an application, the authors in~\cite{lloyd2013quantum} showed how such a density matrix exponentiation method could be used to reveal principle vectors of a given large dataset, with exponential speedup compared to classical algorithms.  

At the same time, the topic of nonlinear science is  fundamental, as many phenomena in nature are inherently nonlinear, such as chaos and nonlinear differential equations. Solving nonlinear systems is generally difficult, and most of them do not admit analytical solutions. Numerical methods typically require substantial resources. It is thus very natural and motivational to explore how quantum computation can tackle such a challenge. Regarding nonlinear algebraic equations, prior works have demonstrated certain improvements in specific regimes. In~\cite{qian2019quantum}, the authors proposed a quantum nonlinear system solver based on Grover's algorithm, which led to a quadratic improvement. While the method is simple, the number of qubits required is proportional to the number of variables, plus many qubits required for precision purposes. Xue et al.~\cite{xue2021quantum} introduced a hybrid classical-quantum method that encompasses the classical Newton method and a quantum tool called $l_{\infty}$ tomography. The advantage of this method turns out not to be provable, as it is only seen via numerics. More recently, Ref.~\cite{xue2022quantum} provided an efficient quantum algorithm for solving quadratic nonlinear systems of equations with an exponential speedup (with respect to the number of variables). While the method seems appealing with such an exponential speedup, it is limited to second order only, i.e., the highest order in each equation is 2. These contributions reveal the inherent difficulty of nonlinear equations, where the speedup (in terms of variables) is usually traded off with other parameters.

Inspired by these developments, we first consider a system of nonlinear algebraic equations where each equation $\{f_i\}$ is a homogeneous polynomial of an even degree over $N$ variables, with generalization of any type of polynomials to be given subsequently. The goal is to find the root or the point at which the values of all functions are zero simultaneously. We remark that multiple roots might exist, and to our knowledge, no method is known to be able to find all of them. While the solution to the linear algebraic equation can be written down explicitly (in the ideal case that no singularity exists), nonlinear algebraic equations do not have an explicit form for the solution. A well-known method to approximately solve such nonlinear algebraic equations is Newton's method. 

\begin{figure}[H]
\centering
\begin{tikzpicture}[>=latex]
\draw[->] (-1,0) -- (6,0) node[below] {$x$};
\draw[->] (0,-1) -- (0,6) node[left] {$y$};

\draw[blue, thick, domain=0:5.8, smooth, variable=\x] plot (\x, {0.5*\x^2 - 3*\x + 4}) node[right] {$f(x) = 0.5x^2-3x+4$};
\draw[red, thick, domain = 4.25:5.8, smooth, variable = \x] plot (\x, {2*\x - 8.5}) node[below right]{}; 
\draw[orange, thick, domain = 0:1.55, smooth, variable = \x] plot (\x, {-2.5*\x +3.875}) node[below]{}; 

\draw[red, dashed] (5,1.5) -- (5,0) node[right] {};
\draw[orange, dashed] (0.5,2.625) -- (0.5,0) node[right] {};

\filldraw[fill= black, draw=black] (5,1.5) circle (2pt) node[right] {$(x_0, f(x_0))$};
\filldraw[fill = black, draw=black] (5,0) circle (2pt) node[below] {$(x_0)$};
\filldraw[fill = black, draw = black] (4.25,0) circle (2pt) node[below] {$(x_1)$};
\filldraw[fill = black, draw = black] (0.5,2.625) circle (2pt) node[right] {$(x_m,f(x_m))$};
\filldraw[fill = black, draw = black] (1.55,0) circle (2pt) node[below] {$(x_n)$};
\filldraw[fill = black, draw =black] (0.5,0) circle (2pt) node[below] {$(x_m)$};
\end{tikzpicture}
    \caption{Illustration of Newton's method in 1 variable case, aiming to find the root of the nonlinear equation $f(x) = 0.5x^2 - 3x +4$. First we initialize the guessed solution $x_0$. The tangent line to $f(x)$ at $x_0$ intersects the $x$-axis at the point $x_1$, which is the updated solution. The procedure is then iterated multiple times until the real root is found, or at least a desired approximation is reached. We note that since the equation might have more than 1 solution, the initialization is critical as the method would drive toward the closest root. On the same figure, we can see that if initially we begin at $x_m$ (on the left), then we end up with different root.   }
    \label{fig:enter-label}
\end{figure}
Roughly speaking, the method begins with a randomly guessed solution. In an iterative manner, we then need to evaluate the so-called Jacobian, which is composed of the gradients of all function $\{f_i\}$ at the given point, giving rise to a system of linear equations. The temporal solution is then updated by subtracting the temporal solution from the solution of the linear equation, and the procedure is then iterated multiple times to drive the initial guess to the  solution. In the case of one variable, the Jacobian is simply the first derivative. Figure~\ref{fig:enter-label} illustrates how Newton's method works, e.g., moving toward the root based on the tangent line. It is well-known that Newton's method is highly effective when the initially randomized guessed root is within the vicinity of the actual root. Our algorithm to be described below is built upon such a Newton iterative produce, similar to that of~\cite{xue2021quantum}. The difference is that our method is fully quantum. Further, we shall see that the particular algebraic form, e.g., homogeneous polynomials of even degree, allows the gradient of those functions to be computed in a closed form, which is more convenient to compute the Jacobian and execute the corresponding Newton method. We remark that this kind of polynomial has also appeared in the context of gradient descent~\cite{rebentrost2019quantum}, where the authors considered the problem of finding local minima of a given objective function. Subsequently, along the way that we generalize our framework to polynomials of an arbitrary degree, we also provide a generalized version of such a gradient descent method proposed in~\cite{rebentrost2019quantum}. 

To highlight our contribution, we emphasize that the number of qubits required, and the running time of our method is logarithmic and polylogarithmic, respectively, in relative to the number of variables in the algebraic equations. Further, our method could be applied to polynomial of any orders. Altogether, it implies the major improvement over previous attempts \cite{xue2021quantum,xue2022quantum, qian2019quantum}, which is enabled by the recent striking quantum singular value transformation framework.

The structure of the paper is as follows. In Section~\ref{sec: preliminaries} we summarize a few key recipes that serve as the building blocks of our main algorithm. We then provide an overview of our objective in Section~\ref{sec: prodes}, which includes the main problem statement and the classical Newton method for solving nonlinear algebraic equations, which eventually boils down to solve a linear equation. Next, we show how to invert the so-called Jacobian matrix in Section~\ref{sec: compgrad} and prepare the right-hand side operator of the linear equation in Section~\ref{sec: preparing}. We then combine the main components and construct our main framework, which is quantum Newton method in Section~\ref{sec: quantumnewtonmethod}. In Section~\ref{sec: qadvantage}, we discuss various critical aspects of our algorithm, which includes the quantum advantage,  and a quantum heuristic method for better initialization and generalization to inhomogeneous polynomials. Such generalization leverages our method for solving nonlinear equations of arbitrary polynomial types, and as a result, provides a completion for these problems.  on quantum gradient descent method in~\cite{rebentrost2019quantum} and~\cite{nghiem2023improved}, e.g, we can execute quantum gradient descent method with any polynomial type.

\section{Preliminaries}
\label{sec: preliminaries}
Here, we summarize the main recipe of our quantum algorithms to be described below. We keep the statements brief and precise for simplicity, with their proofs/ constructions referred to their original works.

\begin{definition}[Block Encoding Unitary]~\cite{low2017optimal, low2019hamiltonian, gilyen2019quantum}
\label{def: blockencode} 
Let $A$ be some Hermitian matrix of size $N \times N$ whose matrix norm $|A| < 1$. Let a unitary $U$ have the following form:
\begin{align*}
    U = \begin{pmatrix}
       A & \cdot \\
       \cdot & \cdot \\
    \end{pmatrix}.
\end{align*}
Then $U$ is said to be an exact block encoding of matrix $A$. Equivalently, we can write:
\begin{align*}
    U = \ket{ \bf{0}}\bra{ \bf{0}} \otimes A + \cdots,
\end{align*}
where $\ket{\bf 0}$ refers to the ancilla system required for the block encoding purpose. In the case where the $U$ has the form 
$$ U  =  \ket{ \bf{0}}\bra{ \bf{0}} \otimes \Tilde{A} + \cdots, $$
where $|| \Tilde{A} - A || \leq \epsilon$ (with $||.||$ being the matrix norm), then $U$ is said to be an $\epsilon$-approximated block encoding of $A$.
\end{definition}

\begin{lemma}[\cite{gilyen2019quantum}]
\label{lemma: improveddme}
Let $\rho = \Tr_A \ket{\Phi}\bra{\Phi}$, where $\rho \in \mathbb{H}_B$, $\ket{\Phi} \in  \mathbb{H}_A \otimes \mathbb{H}_B$. Given a unitary $U$ that generates $\ket{\Phi}$ from $\ket{\bf 0}_A \otimes \ket{\bf 0}_B$, then there exists an efficient procedure that constructs an exact unitary block encoding of $\rho$.
\end{lemma}
The proof of the above lemma is given in \cite{gilyen2019quantum} (see their Lemma 45). \\

\begin{lemma}[Block Encoding of Product of Two Matrices]
\label{lemma: product}
    Given the unitary block encoding of two matrices $A_1$ and $A_2$, then there exists an efficient procedure that constructs a unitary block encoding of $A_1 A_2$.
\end{lemma}
The proof of the above lemma is also given in \cite{gilyen2019quantum}.  \\

\begin{lemma}[\cite{camps2020approximate}]
\label{lemma: tensorproduct}
    Given the unitary block encoding $\{U_i\}_{i=1}^m$ of multiple operators $\{M_i\}_{i=1}^m$ (assumed to be exact encoding), then, there is a procedure that produces the unitary block encoding operator of $\bigotimes_{i=1}^m M_i$, which requires a single use of each $\{U_i\}_{i=1}^m$ and $\mathcal{O}(1)$ SWAP gates. 
\end{lemma}
The above lemma is a result in \cite{camps2020approximate}. 
\begin{lemma}
\label{lemma: As}
    Given oracle access to $s$-sparse matrix $A$ of dimension $n\times n$, then an $\epsilon$-approximated unitary block encoding of $A/s$ could be prepared with gate/time complexity $\mathcal{O}(\log n + \log^{2.5}(\frac{1}{\epsilon}))$.
\end{lemma}
This is also presented in~\cite{gilyen2019quantum}. In~\cite{childs2017lecture}, one can also find a similar construction. 
\begin{lemma}
    Given unitary block encoding of multiple operators $\{M_i\}_{i=1}^m$. Then, there is a procedure that produces a unitary block encoding operator of $\sum_{i=1}^m \pm M_i/m $ in complexity $\mathcal{O}(m)$.
    \label{lemma: sumencoding}
\end{lemma}

\section{Problem Description}
\label{sec: prodes}
We consider a system of nonlinear algebraic equations with $N$ variables and aim to find its root: 
\begin{align}
    f_1( x) = 0, 
    f_2( x ) = 0,
    \dots,
    f_N (x) = 0, 
    \label{eqn: systemnonlinear}
\end{align}
where $x = (x_1,x_2,\cdots, x_N) \in R^N$ and for all $i$, the function $f_i: R^N \rightarrow R$ is some homogeneous polynomial of even degree $2p$ ($p \in \mathbb{Z})$. As mentioned in~\cite{rebentrost2019quantum}, such a polynomial admits an algebraic form: 
\begin{align}
    f_i = \frac{1}{2} x^{\otimes p} A_i x^{\otimes p},
\end{align}
where $A_i$ is a matrix with bounded norm, and of dimension $N^p \times N^p$. It was further mentioned in~\cite{rebentrost2019quantum} that $A_i$ could be decomposed as:
\begin{align}
\label{eqn: ai}
    A_i = \sum_{\alpha=1}^K A_1^{\alpha,i} \otimes \cdots \otimes A_p^{\alpha,i},
\end{align}
where $K$ is just some number that is needed to specify $A_i$. We remark that the above tensor decomposition of $A_i$ is not necessarily known. Such formulation is simply for the gradient computation purpose, as we will see later. In fact, as also mentioned in~\cite{rebentrost2019quantum}, the black box access to entries of $A_i$ (for all $i$) is sufficient for computing the gradient. In the following, we work with such an assumption.  \\

The nonlinear algebraic system~(\ref{eqn: systemnonlinear}) can be solved using the well-known Newton's method, which proceeds iteratively as follows: \\

$\bullet$ We first randomize a vector $x_0$. \\

$\bullet$ Suppose at the $k$-th step, we arrive at some solution $x_k$. We solve the following linear equations:
\begin{align}
    J(x_k) \cdot \Delta_k = F(x_k),
    \label{eqn: LE}
\end{align}
where $F = [f_1,f_2, ..., f_N]^T$,  $F(x_k)$ is the vector-valued function $F$ evaluated at $x_k$, and $J$ is the Jacobian of $F$ evaluated at $x_k$. To be specific, the $i$-th row of $J$, denoted as $J^i$, is the gradient of $f_i$:

$$ J^i =  \Big[ \frac{\partial f_i}{\partial x_1}, \frac{\partial f_i}{x_2}, \dots , \frac{\partial f_i}{\partial x_n} \Big].  $$

$\bullet$ After solving for $\Delta_k$, we update $x_k$: 
$$ x_{k+1} = x_k - \Delta_k. $$

Despite having such a simple look, the above process is not too easy to compute, especially in the quantum context. In order to solve the linear system, one first needs to compute the gradient $ \bigtriangledown f_i$ of all functions to obtain the Jacobian matrix $J$. The next step is to quantumly invert such a Jacobian, which is apparently not symmetric. Furthermore, we note that in the original quantum linear solver~\cite{harrow2009quantum} and its improved version~\cite{childs2017quantum}, the black box access to entries of the matrix of interest is given. However, in this case, we do not have such access. The next step is to prepare the state corresponding to the right-hand side of equation~(\ref{eqn: LE}) and solve the linear system before updating the solution as the final step. \\

In the following sections, we handle these challenges respectively and eventually combine them to execute Newton's method in a quantum mechanical way. \\

\section{Block Encoding the Jacobian J and its inversion}
\label{sec: compgrad}
In order to apply the technique of quantum singular value transformation~\cite{low2017optimal,low2019hamiltonian, gilyen2019quantum} for matrix inversion, one first needs to block encode Jacobian $J$, which consists of gradients of each function $f_i$. Regarding the gradient, as worked out in~\cite{rebentrost2019quantum}, the particular form (homogenous polynomial of an even degree) of $\{f_i\}_{i=1}^N$ allows their gradient (evaluated at some $x$) to be computed as: 
\begin{align}
   \bigtriangledown f_i (x) = D_i(x)\cdot  x,
    \label{eqn: grad}
\end{align}
where
\begin{align}
    D_i(x) = \sum_{\alpha=1}^K\sum_{m=1}^p \left( \prod_{n=1, n\neq m}^p  x^T A_n^{\alpha,i}x \right) A_m^{\alpha,i}.
    \label{eqn: D}
\end{align}
We emphasize that throughout this paper, we use $\Vec{\bigtriangledown} f_i (x)$ and $D_i(x) \cdot x \equiv D_i x$ interchangeably since they represent the same object.   \\

According to~\cite{rebentrost2019quantum}, the gradient operator $D_i$ can be represented as:
\begin{align}
    D_i(x) = \Tr_{1... p-1} \{ (x x^T)^{\otimes p-1} \otimes I) M_D^i\},
\end{align}
where each operator $M_D^i$ is given by:
\begin{align}
    M_D^i = \sum_{\alpha=1}^K \sum_{m=1}^p \left( \bigotimes_{n=1, n\neq m}^p A_n^{\alpha,i} \right) \bigotimes A_m^{\alpha,i}.
\end{align}
Each $M_D^i$ can also be equivalently written as:
\begin{align*}
    M_D^i  = \sum_{m=1}^p M_{D,m}^i,
\end{align*}
 where each $M_{D,m}^i$ is the permutation of matrix $A_i$, i.e., swapping the order of any two composing tensor components in~(\ref{eqn: ai}). As constructed in~\cite{rebentrost2019quantum}, the oracle access to entries of $A_i$ allows access to the above matrix $M_D^i$, e.g., via the following relation:
\begin{align}
    M_D^i = \sum_{j=1}^p Q_j A_i Q_j,
\end{align}
where each $Q_j$ is a permutation matrix that is entry-computable and can be efficiently simulated using $\mathcal{O}(\log (N))$ SWAP gates (see further Ref.~\cite{rebentrost2019quantum}). The first recipe we need is the following, and its proof will be given subsequently. 
\begin{lemma}
\label{lemma: md}
    Given the oracle access to entries of $A_i$ for all $i$, then it is possible to construct the $\epsilon$-approximated block encoding of $M/(ps)$, where $s$ is the sparsity of $M$ and $M$ is the following matrix
    $$ M = \begin{pmatrix}
        M_D^1 & \cdot & \cdot & \cdot \\
        \cdot & M_D^2 & \cdot & \cdot \\
        \cdot & \cdot & \dots & \cdot \\
        \cdot & \cdot & \cdot & M_D^n \\
    \end{pmatrix}, $$
    with time complexity being 
    $$ \mathcal{O}\Big( p(\log(n) + \log^{2.5}(\frac{1}{\epsilon} ) \Big). $$
\end{lemma}

\textit{Proof:} For each $i$, let $M_{D,j}^i\equiv Q_jA_i Q_j $. We consider a larger matrix, which contains all matrices of consideration:
\begin{align*}
    M_{D,j} = \begin{pmatrix}
        M_{D,j}^1 & \cdot & \cdot & \cdot \\
        \cdot & M_{D,j}^2 & \cdot & \cdot \\
        \cdot & \cdot & \dots & \cdot \\
        \cdot & \cdot & \cdot & M_{D,j}^n \\
    \end{pmatrix}.
\end{align*}
As we mentioned, the entries of each $M_{D,j}^i$ can be obtained via the oracle access to entries of $A_i$ for all $i$, e.g., by using Lemma~\ref{lemma: As} to construct block encoding of $A_i/s$ and Lemma~\ref{lemma: product} to construct the block encoding of $Q_j A_i Q_j$. Therefore, entries of such $M_{D,j}$ can be accessed via the same procedure. Lemma~\ref{lemma: As} allows us to prepare the $\epsilon$-approximated block encoding of $M_{D,j}/s$ (for all $j$), where $s$ is the sparsity. Lemma~\ref{lemma: sumencoding} then allows us to prepare the $\epsilon$-approximated block encoding of $M/(ps)$ with the stated time complexity.

For any $i$ in the above sum, we have the following:
\begin{align}
\label{eqn: 16}
    ((x x^T)^{\otimes p-1} \otimes I ) M_D^i (x x^T)^{\otimes p}  &= \sum_{\alpha=1}^K\sum_{m=1}^p  \Big(  \bigotimes_{n=1,n\neq m}^p (x x^T) A_{n}^{\alpha} (xx^T) \Big) \bigotimes A_m^{\alpha} (x x^T) \\
    &= \sum_{\alpha=1}^K\sum_{m=1}^p  \Big(  \bigotimes_{n=1,n\neq m}^p xx^T A_{n}^{\alpha}x  (x x^T) \Big) \bigotimes A_m^{\alpha} (x x^T) \\
    &= (x x^T)^{\otimes p-1} \otimes D_i(x) (x x^T)  \\
    &= (x x^T)^{\otimes p-1} \otimes \bigtriangledown f_i(x) x^T,
\end{align}
where the last line is derived from the fact that $D(x)x \equiv \bigtriangledown f(x)$. Let $I_n$ denote the identity matrix of dimension $n \times n$, and  note that $M$ can be written as:
\begin{align*}
    M = \sum_{i=1}^n \ket{i}\bra{i}\otimes M_D^i.
\end{align*} Then we go through the following manipulation:
\begin{align}
    P &= ( I_n \otimes ((x x^T)^{\otimes p-1} \otimes I ) ) \cdot M \cdot ( I_n \otimes (x x^T)^{\otimes p}  )  \\
    &= ( I_n \otimes ((x x^T)^{\otimes p-1} \otimes I ) ) \cdot (\sum_{i=1}^n \ket{i}\bra{i}\otimes M_D^i ) \cdot ( I_n \otimes (x x^T)^{\otimes p}  ) \\
    &= \sum_{i=1}^n \ket{i}\bra{i} \otimes ((x x^T)^{\otimes p-1} \otimes I ) M_D^i (x x^T)^{\otimes p} \\
    &= \sum_{i=1}^n \ket{i}\bra{i} \otimes \Big( (x x^T)^{\otimes p-1} \otimes \bigtriangledown f_i(x) x^T   \Big). 
\end{align}

Now, suppose, for the moment, that we are presented with a block encoding of $x x^T$. Let $\mathcal{T}$ be the corresponding time complexity of producing $x x^T$. Lemma~\ref{lemma: tensorproduct} allows us to construct the block encoding of $I_n \otimes (xx^T)^{\otimes p-1} \otimes I$ and $I_n \otimes (x x^T)^{\otimes p}$. The block encoding of $M/(ps)$ was presented previously. Then, Lemma~\ref{lemma: product} allows us to prepare the block encoding of $P/(ps)$. Denote this unitary as $U$, for simplicity.

The goal now is to produce the block encoding of the Jacobian $J$. The main tool has been rigorously derived and analyzed in the previous work~\cite{nghiem2023improved}. Therefore, for convenience, we state the following lemma and leave its full discussion in the appendix~\ref{sec: proof}.
\begin{lemma}
\label{lemma: 7}
Given the block encoding of $P/(ps)$ as discussed above, then it is possible to construct the block encoding of the following operator:
\begin{align}
  \hat{O}_J\equiv  \frac{ \gamma^{2p-1} J }{\sqrt{n}ps}, 
\end{align}
where $\gamma < 1$ is some bounded number, and the complexity of the construction is $\mathcal{O}(1)$.
\end{lemma}
We remark that the complexity $\mathcal{O}(1)$ appeared above does not include the complexity of preparing the block encoding of $P/(ps)$, and the value of $\gamma$, while being bounded, changes over different iteration steps. Furthermore, the factor 
${\gamma^{2p-1}}/({\sqrt{ n} ps}) $
in the above formula is seemingly unfavorable. Fortunately, we can remove the factor $ps$ using the technique called amplification~\cite{gilyen2019quantum} (Theorem 30 of \cite{gilyen2019quantum}). Such a method requires the usage of the unitary $U$ that encodes the operator $\hat{O}_J$ $\mathcal{O}( ps \log(ps/\epsilon))$ times. Therefore, the total time required to prepare the $\epsilon$-approximated block encoding of $\gamma^{2p-1}J/\sqrt{n}$ is 
\begin{align}
    \mathcal{O}\Big( ps \log(\frac{ps}{\epsilon}) \ ( \mathcal{T} +  p (\log(n) + \log^{2.5} (\frac{1}{\epsilon}) ) ) \Big).
    \label{eqn: J}
\end{align}
One may wonder why we keep the factor $\gamma^{2p-1}/\sqrt{n}$, which will become clear later when we construct the quantum Newton's method in Section~\ref{sec: quantumnewtonmethod}. 

In a series of work~\cite{chakraborty2018power, chia2020quantum}, the authors showed that given a block encoding of some matrix, say, $\gamma^{2p-1}J/\sqrt{n}$  in our case, it is possible to construct the pseudoinverse $J^{-1}$ of $J$ (we are ignoring the factor $\gamma^{2p-1}/\sqrt{n}$). The core ideas come from the original framework~\cite{gilyen2019quantum}, where the ability to block-encode some matrix translates into the ability to manipulate its singular values. Then, the same method as in Ref.~\cite{childs2017quantum} is executed to achieve the goal of inverting a matrix. We have the following lemma:
\begin{lemma}
\label{lemma: inverseJ}
    Given the block encoding of $\gamma^{2p-1} J /\sqrt{n}$ as described above, the $\epsilon$-approximated block encoding of its pseudoinverse $(\sqrt{n}/\gamma^{2p-1}) \ J^{-1}/\sigma$ is achieved in time complexity 
    $$ \mathcal{O}\left( ps\log(\frac{ps}{\epsilon}) \Big( \mathcal{T} + p \big(\log(n) + \log^{2.5} (\frac{1}{\epsilon})\big) \Big) \cdot  \frac{1}{\sigma} \cdot  polylog(\frac{1}{\sigma\epsilon}) \right),  $$
    where $s$ is the sparsity as defined, and $\sigma$ is some constant that is at least greater than the smallest singular value of $\gamma^{2p-1} J /\sqrt{n}$.
\end{lemma}

It is remarked that the factor $\sigma$ is to guarantee the norm (commonly, $l_2$ norm or operator norm, but the specific choice is not critical) of the inverted matrix, $J^{-1}/\sigma$ is smaller than unity, as typically introduced in the matrix inversion~\cite{harrow2009quantum, childs2017quantum}. In the same context, the problem setup assumes that such a value, $\sigma$ (or more like the conditional number of the matrix), is known for a given matrix of consideration. In our case, the tricky point is that the Jacobian matrix $J$ is changed after each iteration, and hence, the value of $\sigma$ changes. In general, while we know that its spectrum is upper bounded by 1 (see Appendix~\ref{sec: boundJ}, as we can always rescale the matrix to achieve such a condition), the lower bound is unclear. The obvious solution is to find it, and fortunately, this is not a hard problem. Finding the minimum singular/eigenvalue of a given matrix is a very common problem in multiple contexts. For instance, the minimum eigenvalue of a Hamiltonian is the ground state energy.  Recently, in~\cite{nghiem2022quantum}, we have proposed a simple and highly efficient method for finding the largest/ smallest eigenvalues (in magnitude) of a block-encoded matrix.   We simply quote the result here before discussing how to apply it to our problem. 
\begin{lemma}[\cite{nghiem2023improved}]
\label{lemma: 9}
    Given the block encoding with complexity $T_A$ of some positive and Hermitian matrix $A$ of dimension $n \times n$. Then its smallest eigenvalue (and also largest eigenvalue) could be estimated up to an additive error $\epsilon$ in time
    \begin{align}
    \mathcal{O}\Big( \big(\log( \frac{1}{\epsilon}) + \frac{\log(n)}{2}\big)  T_A    \frac{1}{\epsilon}    \Big). 
    \end{align}
\end{lemma}
The above result can only be applied to a positive and Hermitian matrix, i.e., having positive and real eigenvalues. In general, the Jacobian matrix admits negative singular values. In order to apply the method, we need to add the following extra step. Given the block encoding of $J$, its transpose simply contains the block encoding of $J^\dagger$. Lemma~\ref{lemma: product} allows us to construct the block encoding of $J^\dagger J$, whose spectrum is simply the square of singular values of $J$, which are positive. It is easy to see that $J^\dagger J$ is Hermitian. Therefore, we can apply the above lemma to find the smallest eigenvalue of $J^\dagger J$, which is the square of the smallest (in magnitude) singular value of $J$. 

Another important point is that, per Eqn.~(\ref{eqn: J}), the running time of the above Lemma~\ref{lemma: 9} is dominant by the time required to prepare the block encoding of $J$. The error in the above lemma is the additive error resulting from the estimation of the singular value, which is not similar to the $\epsilon$ error that we have used to denote the approximation of the block encoding operator. Therefore, this error term will not be accumulated into the final running time, as the error from this step is independent.

\section{Preparing Right-Hand Side State}
\label{sec: preparing}
Now, we aim to quantumly prepare the right-hand side of the linear equation of interest Eqn.~(\ref{eqn: LE}), namely, the vector 
\begin{align}
    F(x) = \begin{pmatrix}
        f_1 (x) \\
        f_2 (x) \\
        \vdots  \\
        f_n (x) 
    \end{pmatrix},
\end{align}
and for each $i = 1,2,..., n$:
\begin{align}
    f_i(x) = \frac{1}{2} {x^T} ^{\otimes p} A_i x^{\otimes p} .
\end{align}
To deal with $A$'s, we use the following lemma:
\begin{lemma}
\label{lemma: encodingA}
    Given oracle access to entries of $\{A_i\}_{i=1}^n$ then it is possible to prepare the $\epsilon$-approximated block encoding of $A/s$ ($s$ is the sparsity of $A$), where $A$ is the following
    \begin{align*}
    A = \begin{pmatrix}
        A_1/2 & \cdot & \cdot & \cdot \\
        \cdot & A_2/2 & \cdot & \cdot \\
        \cdot & \cdot & \dots & \cdot \\
        \cdot & \cdot & \cdot & A_n/2 \\
    \end{pmatrix},
      \end{align*}
      in time complexity $\mathcal{O}( p\log(n) + \log^{2.5}(1/\epsilon))$.
\end{lemma}
This Lemma can be easily proved. The oracle access to entries of $\{A_i\}_{i=1}^n$, up to a trivial rescale by a factor of $2$, naturally yields entries access to $A$ by using Lemma~\ref{lemma: As}. We note the following representation of $A$, up to a factor of 2, 
\begin{align}
    A =  \frac{1}{2} \sum_{i=1}^n \ket{i}\bra{i}\otimes A_i. 
\end{align}

For now, suppose that we are presented with some block encoding of $x x^T$, where $x$ is some $n$-dimensional vector. Since the block encoding of identity matrix $I$ of  arbitrary dimension is easily prepared, Lemma~\ref{lemma: tensorproduct} allows the construction of 
   $I \otimes  (xx^T)^{\otimes p}$.
Using the block encoding of $A$ (see Lemma \ref{lemma: encodingA}), as well as  the above operator and Lemma~\ref{lemma: product}, then allows us to construct the block encoding of 
\begin{align}
     (I \otimes  (xx^T)^{\otimes p}) A ( I \otimes  (xx^T)^{\otimes p} ).
\end{align}
From this, we perform the following steps:
\begin{align}
  &   (I \otimes  (xx^T)^{\otimes p}) A   (I \otimes  (xx^T)^{\otimes p}) \\
  &= (I \otimes  (xx^T)^{\otimes p})( \frac{1}{2} \sum_{i=1}^n \ket{i}\bra{i}\otimes A_i  ) (I \otimes  (xx^T)^{\otimes p}) \\
  &=  \frac{1}{2} \sum_{i=1}^n \ket{i}\bra{i} \otimes (xx^T )^{\otimes p} A_i (xx^T)^{\otimes p} \\
  &= \sum_{i=1}^n \ket{i}\bra{i} \otimes f_i(x) (xx^T)^{\otimes p} \\
  &= \sum_{i=1}^n f_i(x) \ket{i} \bra{i} \otimes (xx^T)^{\otimes p}.
\end{align}

In the same appendix~\ref{sec: proof} of Lemma~\ref{lemma: 7},  using the same technique, we can show the following:
\begin{lemma}
\label{lemma: 10}
    Given some $\epsilon$-approximated block encoding of $  \sum_{i=1}^n f_i(x) \ket{i} \bra{i} \otimes (xx^T)^{\otimes p}$. Then, it is possible to construct the $\epsilon$-approximated block encoding of 
    $$ \frac{\gamma^{2p-1} F(x) x^T  }{\sqrt{n}} $$
    in time complexity $\mathcal{O}(1)$, where $\gamma$ is some bounded constant.
\end{lemma}
We mention that it is straightforward to obtain the block encoding of $x F(x)^T$ (again, we are ignoring the factor $\gamma^{2p-1}/\sqrt{n}$) by simply taking the transpose of block encoding of $F(x) x^T$. The block encoding of $x F(x)^T$ and $F(x) x^T$, as well as Lemma~\ref{lemma: product}, allows us to construct the block encoding of their product
$$ F(x) x^T x F(x)^T = |x|^2 F(x) F(x)^T.  $$

Using the amplification technique (essentially the uniform singular value transformation, Theorem 30 of \cite{gilyen2019quantum}), one can remove the factor $|x|^2$ to obtain an $\epsilon$-approximation block encoding of $F(x) F(x)^T$ with a further time complexity $\mathcal{O}(\log(|x|)/(|x|^2\epsilon) )$. The remaining question now is, in order to remove such a factor $|x|^2$, we need to know the value. This is exactly the norm of the solution, and it can be estimated using Lemma~\ref{lemma: 9}. To show this, we simply make use of the following observation: at a given iteration step, we are presented with the block encoding of $x x^T$. This operator has a special property that $|x|^2$ is its largest eigenvalue. Therefore, Lemma~\ref{lemma: 9} is sufficient to estimate such factor $|x|^2$.

\section{Quantum Newton Method}
\label{sec: quantumnewtonmethod}
To recap what we have obtained, assuming that we have multiple block encodings of $x x^T$, we have the following recipes so far: \\

$\bullet$ \ $\epsilon$-approximated block encoding of $(\sqrt{n}/\gamma^{2p-1}) J^{-1}/\sigma$, where $\sigma$ is the smallest singular value of $J$. We note that the value of $\sigma$ depends on $J$ at each iteration, and it can be estimated using Lemma~\ref{lemma: 9}. We remark that this is the most time-consuming step compared to preparing the right-hand side state in the previous section. Therefore, the running time of our algorithm is dominant by the inversion of $(\sqrt{n}/\gamma^{2p-1})  J^{-1}/\sigma$. \\

$\bullet $ \  $\epsilon$-approximated block encoding of $F(x) x^T$, $x F(x)^T$ and $F(x) F(x)^T$. The time complexity of these block encodings is (asymptotically) less than the time required to prepare $J^{-1}/\sigma$ as mentioned above. Lemma~\ref{lemma: product} is then used to construct the $\epsilon$-approximated block encoding of $J^{-1} F(x) x^T /\sigma, x (F(x) J^{-1})^T/ \sigma$ and  
$J^{-1} F(x) (J^{-1} F(x) )^T / \sigma^2$. 

Now, we are ready to present our main quantum Newton's method. We first remind the readers that from the linear equation~(\ref{eqn: LE}), at the $k$-th iteration step, we have $\Delta_k = J^{-1} F(x_k)$. We then update the solution as the following:
\begin{align}
    x_{k+1}  &= x_k - \Delta_k = x_k - J^{-1} F(x_k). 
\end{align}
The `density matrix' version of the above solution is:
\begin{align}
    x_{k+1} x_{k+1}^T &= (x_k - J^{-1} F(x_k) ) ( x_k - J^{-1} F(x_k) )^T \\
    &= x_k x_k^T - x_k (J^{-1} F(x_k))^T - J^{-1} F(x_k) x_k^T + J^{-1} F(x_k) ( J^{-1} F(x_k))^T.  
\end{align}
Before heading to the final stage, we emphasize an important point that accounts for the reason why we keep the factor $\gamma^{2p-1}/\sqrt{n}$ throughout the work. The above formula contains the term $J^{-1} F(x)x$ (where $x$ refers to any input vector). The inversion of $\gamma^{2p-1} J/\sqrt{n}$ is $\sqrt{n} J^{-1}/\gamma^{2p-1}$, when it multiplies with the block encoding of $\gamma^{2p-1} F(x)x^T / \sqrt{n}$, the unwanted factors will cancel out. Therefore, we do not need to use the amplification technique to remove the factor before executing the algorithm, which results in a very efficient procedure. 

We have already outlined steps to obtain the block encoding of the last three terms in the above equation divided by $\sigma_k$. Given the block encoding of $x_k x_k^T$ (e.g., from the previous iteration step), it is straightforward to obtain the block encoding of $x_k x_k^T /\sigma$. Then Lemma~\ref{lemma: sumencoding} can be used to obtain the block encoding of 
$$ \frac{x_k x_k^T - x_k (J^{-1} F(x_k))^T - J^{-1} F(x_k) x_k^T + J^{-1} F(x_k) ( J^{-1} F(x_k))^T  }{4 \sigma_k} = \frac{x_{k+1}x_{k+1}^T}{4\sigma_k}. $$
Therefore, once we obtain the block encoding of the above operator, we produce the block encoding of $x_{k+1}x_{k+1}^T$ with a further $\mathcal{O}(\sigma_k \log(\sigma_k))$ time complexity \cite{gilyen2019quantum}. \\

To be more specific, our quantum Newton's method, at some $k$-th iteration step with the input as the block encoding of $x_k x_k^T$, proceeds as summarized below: \\

$\bullet$ Estimate the norm $|x_k|^2$. \\

$\bullet$ Construct the $\epsilon$-approximated block encoding of Jacobian $\gamma_k^{2p-1}J_k / \sqrt{n}$. \\

$\bullet$ Estimate the lowest singular value (in magnitude), $\sigma_k$, of $\gamma_k^{2p-1}J_k / \sqrt{n}$. \\

$\bullet$ Construct the approximated block encoding of $\sqrt{n} J_k^{-1}/ \gamma_k^{2p-1}\sigma_k$. \\

$\bullet$ Construct the approximated block encoding of $\gamma_k^{2p-1} F(x_k) x_k^T/\sqrt{n}$ and  $\gamma_k^{2p-1} x_k F(x_k)^T / \sqrt{n}$.  \\

$\bullet$ Produce the approximated block encoding of $x_{k+1} x_{k+1}^T$. \\

Now, we analyze the running time. Let $\mathcal{T}_k$ denote the time required to produce the block encoding of $x_k x_k^T$. From the list above, the time required to produce the block encoding of $x_{k+1} x_{k+1}^T$ is 
\begin{align*}
    \mathcal{O} \Big( ps \log(\frac{ps}{\epsilon} ) \big(\log(n) + \frac{1}{\sigma_k} {\rm polylog}( \frac{1}{\sigma_k \epsilon}  ) \big)\big( \mathcal{T}_k + \log(n) + \log^{2.5} (\frac{1}{\epsilon})\big)  \Big). 
\end{align*}
Recall that with the block encoding of $x_k x_k^T$, the same algorithmic procedure produces the block encoding of $x_{k+1} x_{k+1}^T$. Therefore, one can find the above running time using a recursive formula. We summarize our result in the following theorem.
\begin{theorem}
\label{thm: mainthm}
    Given a nonlinear algebraic system as described in Eqn.~(\ref{eqn: systemnonlinear}) and a fixed total iteration step $T$, then there exists a quantum algorithm that begins with a block encoding of some operator $x_0 x_0^T$ and produces an $\epsilon$-close block encoding of $x_r x_r^T$, where $x_r$ is the approximated root of the given algebraic equation. The running time of the algorithm for a total of $r=T$ iterations is:
    \begin{align*}
        \mathcal{O} \Big( (ps\log(\frac{ps}{\epsilon})))^{T+1} \big( \log(n) + \frac{1}{\sigma_{min}} {\rm polylog}( \frac{1}{\sigma_{min} \epsilon}  ) \big)^{T+1} \big( \log(n) + \log^{2.5} ( \frac{1}{\epsilon} )\big) \Big),
    \end{align*}
    where $\sigma_{min}$ refers to the minimum value among all $\{ \sigma_{k} \}_{k=1}^T$. 
\end{theorem}
The initial operator $x_0 x_0^T$ can be easily prepared by first choosing a random unitary that generates the state $\ket{x_0} \equiv x_0$, and by adding an ancilla $\ket{0}$ and using Lemma~\ref{lemma: improveddme} to prepare the block encoding of $x_0 x_0^T$. In that case, our initial guess lies on the unit sphere as $|x_0| =1$. We can also choose an initial guess that has its norm less than $1$, e.g., $|x_0|^2 < 1$. The way to achieve this was previously discussed in appendix B of Ref.~\cite{nghiem2023improved}, with the general idea being that we aim to encode some new operator $x x^T$ with a dimension greater than $x_0 x_0^T$ and then pay attention only to the top left corner, e.g., the $n \times n$ matrix. \\

\section{Discussion on Quantum Advantage and Generalization}
\label{sec: qadvantage}
\subsection{Classical Algorithm and Quantum Advantage}
The most dominant step for classical Newton's method is to compute the right-hand side of Eqn.~(\ref{eqn: LE}). Recall that for each $i$:
\begin{align}
    f_i = x^{\otimes p} A_i x^{\otimes p},
\end{align}
each $A_i$ is an $s$-sparse matrix of size $n^p \times n^p$. Therefore, computing the right-hand side of Eqn. (\ref{eqn: LE}) takes complexity
  $\mathcal{O} \Big( n^{p+1} s  \Big)$. 
In order to compute the entries of Jacobian $J$, the classical algorithm needs to evaluate the following term
\begin{align*}
    D_i(x) \cdot x = \sum_{\alpha=1}^K\sum_{k=1}^p \Big( \prod_{j=1, j \neq k}^p  x^T A_j^{\alpha,i}x \Big) A_k^{\alpha,i} \cdot x,
\end{align*}
which takes a further complexity $\mathcal{O}( Kp^2 n s )$. Finally, solving the linear equation~(\ref{eqn: LE}) takes a further  $\mathcal{O} ( n^3  ) $ time.
Therefore, summing up the steps, the total time required for a total of $T$ iteration steps is: 
\begin{align}
\label{eq:classicalT}
    \mathcal{O}\Big( T( n^3 +  Kp^2 n s + n^{p+1} s )\Big).
\end{align}
For large $p$, e.g., $p \geq 3$, the task is practically intractable on a classical computer if we manually follow the above steps. It is well-known that per Newton's method, if the initial solution $x_0$ is sufficiently close to the real zero of given equations, then the solution converges quadratically fast, which means that the total number of iteration times $T \approx \mathcal{O}(\log \log(1/\epsilon))$ where $\epsilon$ is the desired error. The above running time indicates that the speedup of the quantum algorithm is superpolynomial, almost exponential, with respect to the number of variables.

We remark that the problem of solving a system of polynomial equations was, in fact, Smale's 17th problem, where he asked if there exists a polynomial-time classical algorithm that outputs an approximate zero of such equations, given that these equations could be computed at any given point on demand. The answer turned out to be positive after multiple efforts, which are summarized and finally settled in the seminal work~\cite{burgisser2011problem}. The best-known running time, as mentioned in~\cite{burgisser2011problem} is $\mathcal{O}(n^{\log \log(n)} )$, which is almost linear in the number of variables $n$. We emphasize an important subtlety, namely, they assume the ability to compute any given polynomial via certain means; however, the structure of these functions, as well as the complexity of how to compute them efficiently is not taken into account. Therefore, the time complexity of our quantum algorithm, as stated in Theorem~\ref{thm: mainthm}, should be compared to the classical one shown in Eqn.~(\ref{eq:classicalT}).  

 Newton's method typically works well given that we randomly initialize a vector within the vicinity of the real solution. Therefore, it is reasonable that Newton's method (both classical and quantum versions) will succeed after a few iterations, e.g., when $T$ is small. In this case, a huge quantum speedup is obtained. Now, we move on to discuss a way to achieve better initialization from a quantum algorithmic perspective.

\subsection{Quantum Approach for Initialization}
Aside from the particular algebraic form that we impose, the behavior of functions of interest is quite critical. We recall that the Newton method is usually effective when the initially randomized solution is closed to the real one, as the method would succeed after after a few iterations. Therefore, it is quite important to sort out a way to prepare such a good initialization. A general strategy to prepare such a thing is not known to us. Here, we aim to propose a way to accelerate such a task, inspired by the way that each of the functions $f_i$ can be computed quickly using a quantum computer. 

Recall that for each $i = 1,2,..., n$:
\begin{align}
    f_i(x) = \frac{1}{2} (x^T)^{\otimes p} A_i (x^{\otimes p} ),
\end{align}
and 
from Lemma \ref{lemma: 10}  we have the block encoding of 
 \begin{align*}
    A = \begin{pmatrix}
        A_1/2 & \cdot & \cdot & \cdot \\
        \cdot & A_2/2 & \cdot & \cdot \\
        \cdot & \cdot & \dots & \cdot \\
        \cdot & \cdot & \cdot & A_n/2 \\
    \end{pmatrix}.
      \end{align*}
Suppose we have block encoding of $xx^T$, then from previous sections, we have the block encoding of $(xx^T)^{\otimes p}$. It is simple to prepare the block encoding of $I_n \otimes (xx^T)^{\otimes p}$ using the standard method introduced above. We note the following useful equality:
\begin{align}
    (xx^T)^{\otimes p} (A_i)/2  (xx^T)^{\otimes p} &= f_i(x) (xx^T)^{\otimes p}.
\end{align}
Then, we use Lemma~\ref{lemma: product} to prepare the block encoding of the following operator:
\begin{align}
    (I_n \otimes (xx^T)^{\otimes p}) \ A  \ (I_n \otimes (xx^T)^{\otimes p}) = \sum_{i=1}^n \ket{i}\bra{i} \otimes f_i(x) (xx^T)^{\otimes p},
\end{align}
which has a matrix form
\begin{align}
    \begin{pmatrix}
        f_1(x) |x|^{2p} \ket{x}\bra{x} & \cdot & \cdot & \cdot \\
        \cdot & f_2(x) |x|^{2p} \ket{x}\bra{x} & \cdot & \cdot \\
        \cdot & \cdot & \dots & \cdot \\
        \cdot & \cdot & \cdot & f_n(x) |x|^{2p} \ket{x}\bra{x} \\
    \end{pmatrix}.
\end{align}
Recall that each operator $\ket{x}\bra{x}$ is a projector, hence having an eigenvalue $1$. Using Lemma~\ref{lemma: 9}, we can estimate the maximum eigenvalue of the above matrix, which is one of the $f_i(x) |x|^{2p}$. Then from the estimated norm $|x|^2$, we are able to find 
$\max_i f_i(x)$
in time $\mathcal{O}(p\log^2 n)$ (where we ignore the error dependence). Using this procedure, we aim to find the vector $x$ so that all $f_i(x)$ are small as desired, e.g., that $x$ is near the real solution. At first glance, this method seems to be quite heuristic as $x$ is, again, picked from random. However, the procedure is done in logarithmic time, which is faster than our quantum Newton's method (see Thm.~\ref{thm: mainthm}) and apparently faster than any classical method. Therefore, it can be used at first to gauge the range of the root before executing the main algorithm. 

\subsection{Generalizing to Arbitrary Polynomial }
\label{sec: generalization}
Here, we pick up a point that has been briefly mentioned in Ref.~\cite{rebentrost2019quantum}, and we generalize our method to any polynomial, or more precisely, monomial, of any degree that is not necessarily homogeneous. Recall from~\cite{rebentrost2019quantum} that an inhomogeneous polynomial can be written as:
\begin{align}
    f(x) = \sum_{j=1}^{p-1} (c_j^T x) \ \prod_{i=1}^{j-1} (x^T B_{ij} x),
\end{align}
where all the vectors $c_j$ and matrices $B_{ij}$ define our polynomial. We consider the term $(c_j^T x) \ \prod_{i=1}^{j-1} (x^T B_{ij} x)$, and observe that the product $\prod_{i=1}^{j-1} \big(x^T B_{ij} x\big) $ is exactly homogeneous polynomial of an even degree. Since the gradient of the sum of functions is the sum of the gradient of all the functions involved,  we now consider the specific term $g(x) = (c^T x) \ \prod_k (x^T B_k x)$. Using chain rule, the partial derivative of $f(x)$ with respect to some variable $x_m$ is:
\begin{align}
 \frac{\partial g(x)}{\partial x_m} = \frac{\partial (c^T x)}{\partial x_m} \Big( \prod_k x^T B_k x \Big) + (c^T x) \frac{\partial \big(\prod_k x^T B_k x\big)}{\partial x_m}.
\end{align}
Therefore, in vector form, we have that:
\begin{align}
    \nabla g(x) =  \Big( \prod_k x^T B_k x \Big) \ c + (c^T x) \nabla \Big(\prod_k x^T B_k x\Big).
\end{align}
Now, we aim to produce something similar to Eqn~(\ref{eqn: 16}) but with a new gradient formula instead, as it will contribute to the Jacobian matrix construction, and the generalization to multiple equations is straightforward. To be more specific, we want to obtain block encoding of $(xx^T)^{\otimes p-1} \otimes \nabla g(x) x^T$, where we have assumed without loss of generality that the homogeneous term in $g(x)$, which is $\prod_k x^T B_k x $, has degree $2p$, which is equivalent to the expression $(x^T)^{\otimes p} B x^{\otimes p} \equiv b(x)$. Further, we assume that vector $c$ is generated by some quantum circuit $U_C$. Using Lemma~\ref{lemma: improveddme}, preparing the block encoding of $cc^T$ is straightforward. 

First, we aim to produce the block encoding of the first part $\big( \prod_k x^T B_k x \big) \ c x^T$. We know that given the access to entries of $B$, it is simple to obtain the block encoding of $B$ (using Lemma~\ref{lemma: As}). From the block encoding of $(xx^T)^{\otimes p} \otimes cc^T$, We use Lemma~\ref{lemma: product} to prepare the block encoding of $(xx^T)^{\otimes p} \otimes cc^T  \  (B \otimes I) \ (xx^T)^{\otimes p} \otimes cc^T = (xx^T)^{\otimes p } \otimes b(x) cc^T$. Now we use SWAP gates to exchange the order: $xx^T \otimes cc^T \longrightarrow xc^T \otimes cx^T$, we then obtain the block encoding of $(xx^T)^{\otimes p-1} \otimes xc^T \otimes b(x) cx^T$.

From a procedure that has resulted in Equation~(\ref{eqn: 16}), we have the block encoding of $(xx^T)^{\otimes p-1} \otimes \nabla b(x) x^T$. Using Lemma \ref{lemma: tensorproduct} we can obtain the block encoding of $(xx^T)^{\otimes p-1} \otimes \nabla b(x) x^T \otimes cc^T$. Using SWAP gates to exchange the position (for convenience): $\nabla b(x) x^T \otimes cc^T \longrightarrow cc^T \otimes \nabla b(x) x^T$. From the block encoding of $xx^T$ and $cc^T$, we use Lemma~\ref{lemma: product} to prepare the block encoding of $xx^T \cdot cc^T = (x^T c) xc^T$. It is trivial to prepare the block encoding of $I_n^{\otimes p-1} \otimes  (x^T c) xc^T \otimes I_n$. Therefore, using Lemma~\ref{lemma: product}, we can prepare the block encoding of the following product 
$$ I_n^{\otimes p-1} \otimes  (x^T c) xc^T \otimes I_n  \cdot (xx^T)^{\otimes -1} \otimes cc^T \otimes \nabla b(x)x^T = (xx^T)^{\otimes p-1} \otimes xc^T \otimes (x^T c) \nabla b(x) x^T.  $$
Lemma \ref{lemma: sumencoding} can be used to obtain the block encoding of $ (1/2) (xx^T)^{\otimes p-1} \otimes xc^T \otimes (b(x)c + (x^T c) \nabla b(x) )x^T = 1/2 \ (xx^T)^{\otimes p-1} \otimes xc^T \otimes \nabla g(x)  x^T$, which has a similar form to Eqn.~\ref{eqn: 16}. The above procedure is done for one function only; however, the extension to multiple functions is carried out in a straightforward manner as the procedure after Eqn.~(\ref{eqn: 16}), which basically encodes the operator $(xx^T)^{\otimes p-1} \otimes \nabla f_i(x) x^T$ to the diagonal block. In this inhomogeneous setting, $(xx^T)^{\otimes p-1} \otimes \nabla f_i(x) x^T$ is simply replaced by $(xx^T)^{\otimes p-1} \otimes xc^T \otimes \nabla g_i(x) x^T$, and, e.g., we will obtain the block encoding of $\sum_{i=1}^n \ket{i}\bra{i}\otimes (xx^T)^{\otimes p-1} \otimes xc^T \otimes \nabla g_i(x) x^T$. Then, a similar procedure to Lemma~\ref{lemma: 7} can be used to obtain the desired Jacobian, albeit the prefactor is different. In order to complete the method, we need to extend Lemma~\ref{lemma: 10} to the case of an inhomogeneous polynomial as well, which is straightforward, as it is just a multiplication of $c^T x$. From Lemma~\ref{lemma: 10}, we have obtained the block encoding of $\sum_{i=1}^n b_i(x) \ket{i}\bra{i} \otimes (xx^T)^{\otimes p}$. We simply need to use the block encoding of $I_n \otimes I_n^{\otimes p-1} \otimes cc^T$ and Lemma~\ref{lemma: product}, plus the fact that $xx^T cc^T = (x^T c) xc^T = (x^T c) xc^T$ (both $x$ and $c$ are real vectors), to produce the block encoding of $\sum_{i=1}^n (c^T x) b_i(x) \ket{i}\bra{i} \otimes (xx^T)^{p-1} \otimes xc^T$. Then, the rest is carried out as we outlined after Lemma~\ref{lemma: 10} to produce the desired ingredients to execute the quantum Neweton's method. 

As a simple but quite striking corollary, the above construction of the gradient operator for arbitrary polynomial (that includes both homogeneous and inhomogeneous terms) yields a completed version of the quantum gradient descent method proposed in~\cite{rebentrost2019quantum} and recently improved in~\cite{nghiem2023improved}, both of which only dealt with homogeneous polynomials of an even degree.

\section{Motivating Examples And Potential Applications/Extensions}
\label{sec: examples}
To motivate the application of our method, here we provide some examples from the physical context that showcase the appearance of nonlinear algebraic equations. 
\subsubsection{Nonlinear Optical Process/ Gross-Pitaevskii Equation}
In the context of nonlinear optics, an envelope of light beam through a certain medium could be described by a nonlinear Schrodinger equation. Full literature could be found in multiple standard texts, such as~\cite{boyd2008nonlinear}. For our purpose of demonstration, in the simplest form, the time-dependent Gross–Pitaevskii equation reads:
\begin{align}
i\frac{\partial \psi}{\partial t} + \frac{\hbar^2}{2m} \frac{\partial^2 \psi}{\partial x^2} + V(x) \psi + g|\psi|^2 \psi = 0,
\end{align}
where all the parameters $m, V(x) ,g$ are corresponding properties of light and medium in the context of nonlinear optics or the atomic mass, external field, and interaction strength in the BEC context. In order to solve the above equation numerically, a discretization called the Crank-Nicolson method could be used. Let $\psi_j^n$ denote the numerical approximation of wave function $\psi(x_j, t_n)$ at spatial grid points $x_j$ and time $t_n$. A discretized version of the above nonlinear Schrodinger equation is then:
\begin{equation}
\label{eqn: discretenonlinear}
\begin{split}
i\frac{\psi_j^{n+1} - \psi_j^n}{\Delta t} + \frac{\hbar^2}{4m}\Bigg(&\frac{\psi_{j+1}^{n+1} - 2\psi_j^{n+1} + \psi_{j-1}^{n+1}}{\Delta x^2} \\
&+ \frac{\psi_{j+1}^n - 2\psi_j^n + \psi_{j-1}^n}{\Delta x^2}\Bigg) + V_j^n\frac{\psi_j^{n+1} + \psi_j^n}{2} + g\frac{|\psi_j^{n+1}|^2 + |\psi_j^n|^2}{2}\psi_j^{n+1} = 0,
\end{split}
\end{equation}
where $\Delta t, \Delta x$ and $V_j^n$ are time step, grid spacing, and potential at the grid point $(x_j, t_n)$. The above equation at different grid points represents a coupled nonlinear algebraic equation, which could be solved using the classical Newton method to obtain a vector solution that contains all $\{ \psi_j^n \}$ at different grid points. However, we point out two issues. First, our method is only applied to a nonlinear system where each equation is a homogeneous polynomial of even degree. Second, the above equation (Eqn.~\ref{eqn: discretenonlinear}) actually contains both linear and nonlinear terms. In order to integrate our method, an adjustment is hence required. Apparently in Section \ref{sec: generalization} we showed how to deal with this scenario, but here we point out that it can be alternatively done in a simpler manner.

Regarding the first issue, which is the degree of polynomial, we consider the following simple example. Suppose we have a nonlinear algebraic system:
\begin{align}
    2x^2y + xyz + z^3 = 0, \\
    xy^2 +  x^3 + y^3 = 0,\\
    x^2z + y^2z + xz^2 = 0,
\end{align}
The above equations contain homogeneous polynomials of odd degree (i.e., 3), which makes it difficult to implement our method. We observe that if we add an extra variable $m$, and consider the following equations instead:
\begin{align}
    2x^2ym + xyzm + z^3 m = 0, \\
    xy^2 m  +  x^3m + y^3m = 0,\\
    x^2zm + y^2zm + xz^2m = 0, \\
    x^2 - m^2= 0.
\end{align}
Then, apparently, the new system has the same $x,y,z$ solution as the old one. One can see that from the fact that the last equation (of the above ones) indicates $x = \pm  m$, and those $x,y,z$ that satisfy the original equations also satisfy the new system. The difference is that the new system has an even degree, so we can apply our method. The extra equation has trivial coefficients and, hence, is accessible in a quantum context. The trick we mention here is quite analogous to the Hermitian embedding trick in linear system context~\cite{harrow2009quantum}, which can be used to deal with non-Hermitian matrices. 

To deal with the second problem, e.g., linear terms, we first observe that suppose a function (of multiple variables) $f$ consists of two components, $f_1$ and $f_2$, e.g., $f = f_1 + f_2$. Then the derivative of $f$ with respect of any variables $x$ is: 
\begin{align}
    \frac{\partial f}{\partial x}  = \frac{\partial f_1}{\partial x} + \frac{\partial f_2}{\partial x},
\end{align}
which means that the gradient vector of $f$ is simply the sum of gradient of $f_1$ and $f_2$. Additionally, if we have a linear function (in each variable), e.g.,
\begin{align}
    g(x,y,z) = ax + by + cz,
\end{align}
then the gradient vector is simply $(a,b,c)^T$. So our system has, say, three functions, $g_1, g_2, g_3$, sharing the same form as above, e.g., 
\begin{align}
    g = 
    \begin{pmatrix}
        g_1 = a_1x + b_1y + c_1 z \\
        g_2  = a_2x + b_2y + c_2 z\\
        g_3  = a_3 x + b_3 y + c_3 z
    \end{pmatrix},
\end{align}
then the Jacobian $J_g$ of the system is simply composed of the linear coefficients:
\begin{align}
    J_g =  \begin{pmatrix}
        a_1 & b_1 & c_1 \\
        a_2 & b_2 & c_2 \\
        a_3 & b_3 & c_3 
    \end{pmatrix}.
\end{align}
Now, we observe that the nonlinear equation in Ref.~\ref{eqn: discretenonlinear} contains both linear and nonlinear terms. Therefore, if we separate the function on the left-hand side of Eqn.~(\ref{eqn: discretenonlinear}) into linear and nonlinear parts, then the Jacobian of such equations, which basically contains gradients of corresponding functions, is the sum of the Jacobians from the constituting functions. The Jacobian of the linear constituent function is easily computed based on the above example. The Jacobian of the nonlinear constituent is computed using the algorithm outlined in our discussion above.  \\

We remark that the above nonlinear Schrodinger equation is essential to understand the phenomena of so-called \textit{solitons}, which are localized wave packets that maintain their shape through propagation within the medium. Therefore, the solution to the above equation, in principle, can be used for further processing/manipulation to investigate the properties of solitons. For example, as presented in~\cite{infeld2000nonlinear}, the amplitude of the soliton is:
\begin{align}
    A = \max |\psi(x,t)|,
\end{align}
which contains information about the energy carried out by the soliton. In our case, the quantum method returns a block encoding of the operator $\ket{\psi} \bra{\psi}$, where $\ket{\psi}$ is a vector that contains the wave function over the grids (both time and space). The solution to finding the maximum entry in such a matrix turns out not to be so simple, as the description of the matrix is not explicit. A quantum method for finding a minimum/maximum item within a database has been done in~\cite{durr1996quantum}; however, similar to Grover's search algorithm, it is a query-based model, which relies on a given black box that computes the function of interest. In our case, we have a matrix of interest block encoded via some unitary. The solution to find the maximum entry given this model is not known up to our knowledge. We point out that recently, there is a work by Childs et al \cite{childs2021quantum} that aims to investigate properties, such as trace, rank, of a matrix given a query that returns its action on some vector. While it is not obvious of how relevant it is to our problem, we note that the block encoding of the matrix of interest could give us a mean to perform certain operations, such as multiplying it with some vector, which might be useful for investigating properties of given matrix. We leave the problem for further research.  

\subsubsection{Lotka-Volterra Equations for Modeling}
In population dynamics, Lotka-Volterra equations~\cite{wangersky1978lotka}  are often used to model the interaction between predator and prey species. In its general form, the equation is:
\begin{align}
    \frac{dV}{dt} = \alpha V - \beta PV, \\
    \frac{dP}{dt} = - \gamma P  + \delta PV,
\end{align}
where $\alpha, \beta, \gamma, \delta$ are parameters representing interaction strengths and growth rates; $V$ and $P$ are population density of prey and predator.  To discretize the above system, using Euler's method with a time step $\Delta t$, we obtain:
\begin{align}
    V_{n+1} = V_n + \Delta t ( \alpha V_n - \beta V_n P_n  ), \\
    P_{n+1} =  P_n + \Delta t ( -\gamma P_n + \delta V_n P_n  ),
\end{align}
which are equivalent to
\begin{align}
    V_{n+1} - V_n  - \Delta t ( \alpha V_n - \beta V_n P_n  ) = 0, \\
    P_{n+1} -  P_n - \Delta t ( -\gamma P_n + \delta V_n P_n  ) = 0.
\end{align}
The above equations at multiple points form a nonlinear algebraic system. Again, they contain both linear and nonlinear terms, which can be handled in the same manner as we have described in the previous paragraphs.

\subsubsection{Intersection Of Algebraic Vaieties}
Algebraic variety is a fundamental object in algebraic geometry (see Figure \ref{fig:main} for some illustrating examples). Formally, it is a basically a set of solutions to a system of polynomial equations over a field. 

\begin{figure}[H]
    \centering
    \begin{subfigure}[b]{0.3\textwidth}
        \centering
        \begin{tikzpicture}
            \draw[->] (-2,0) -- (2,0) node[right] {$x$};
            \draw[->] (0,-2) -- (0,2) node[above] {$y$};
            \foreach \x in {-1,1}
                \foreach \y in {-1,1}
                    \filldraw[red] (\x,\y) circle (1pt);
        \end{tikzpicture}
        \caption{Affine space $\mathbb{A}^2$: basically four points on $x-y$ plane}
        \label{fig: affineA4}
    \end{subfigure}
    \hfill
    \begin{subfigure}[b]{0.3\textwidth}
        \centering
        \begin{tikzpicture}
            \draw[thick,->] (-2.0,0) -- (2.0,0) node[right] {$x$};
            \draw[thick,->] (0,-2.0) -- (0,2.0) node[above] {$y$};
            \draw[color = blue, domain=0:360,samples=100] plot ({1.5*cos(\x)}, {1.5*sin(\x)});
        \end{tikzpicture}
        \caption{Algebraic curve: a unit circle}
        \label{fig:sub2}
    \end{subfigure}
    \hfill
    \begin{subfigure}[b]{0.3\textwidth} 
        \centering
        \begin{tikzpicture}
            \draw[->] (-2,0) -- (2,0) node[below] {$x$};
            \draw[->] (0,-2) -- (0,2) node[left] {$y$};
            \draw[color = orange,domain=-2.0:2.0,samples=100,variable=\x,smooth] plot (\x,{sqrt(\x*\x*\x - 3*\x + 2)});
            \draw[color = orange, domain=-2.0:2.0,samples=100,variable=\x,smooth] plot (\x,{-sqrt(\x*\x*\x - 3*\x + 2)});
        \end{tikzpicture}
        \caption{Algebraic surface: an elliptic curve}
        \label{fig:sub3}
    \end{subfigure}
    \caption{Some example of algebraic varieties. Figure \ref{fig:sub2} features a one-dimensional algebraic variety. Figure \ref{fig:sub3} features two-dimensional algebraic variety.}
    \label{fig:main}
\end{figure}

In algebraic geometry, one of the main theme is studying geometric shapes defined by polynomial equations. Given multiple varieties, then pictorially, one can imagine that their intersection form exactly a system of nonlinear algebraic equation. Solving for such intersection for a high dimensional/ complicated variety is classically difficult, and in fact, as we pointed out in the work, for degree $\geq 3$, solving such equation is even intractable. At the same time, many important application tasks are based on algebraic geometry concepts, in particular, algebraic variety. In robotics and computer-aided design (CAD) \cite{mortenson1997geometric}, shapes are represented using algebraic varieties, and their intersection could model collision. Thus, finding the intersection properly could potentially be used for collision avoidance strategy in robot motion planning. Furthermore, in the area of robot vision and image processing \cite{szeliski2022computer}, algebraic varieties, such as algebraic curves and surfaces are also used for representing object and and scenes in images. Stemming from our efficient quantum algorithm, we envision a future research that designs quantum algorithm for object recognition, feature extraction, etc from given algebraic representation of objects and scenes, as mentioned.

\section{Conclusion}
In this article, we have developed a quantum algorithm for solving nonlinear algebraic equations of arbitrary polynomial type.  By employing well-known quantum tools, such as the unitary block encoding method, an iterative procedure building upon classical Newton's method is constructed to drive the initially randomized guess to the close-to-real solution, which is expected to succeed after only a few iteration steps. Additionally, we have provided two examples from physical, biophysics and algebraic geometry contexts to suggest the potential application of our method, as well as suggesting another way to extend our method to deal with a more specific setting. 

As we have mentioned, our work adds another instance of the quantum effort to the nonlinear regime, continuing the progress in~\cite{xue2022quantum, xue2021quantum,qian2019quantum}. The time complexity of our algorithm is polylogarithmic with respect to the dimension $n$, exponential in the iteration step $T$, and polynomial in the degree $p$. We emphasize that our method can deal with polynomials of arbitrary type, e.g., arbitrary degree, homogeneous, and inhomogeneous type of polynomial, which was not the case in~\cite{xue2022quantum}. meanwhile, the running time is kept polylogarithmic with respect to the number of variables. Furthermore, the number of qubits required in our work is also logarithmic in the number of variables, which is more efficient in running time and hardware resources than the method in~\cite{qian2019quantum}. We remark that progress in nonlinear science is fundamentally important. In~\cite{liu2021efficient}, the authors quantumly tackled the dissipative nonlinear differential equations using the linearization technique and combined them with the well-known quantum method, e.g., the quantum linear solver. The two examples that we provided above were nonlinear partial differential equations, and as we see, discretizing them leads to nonlinear equations. The fact that we attempt to solve them quantumly from a nonlinear perspective paves a new quantum way to deal with the complicated types of differential equations, which is apparently a rich avenue deserving of further efforts. It thus contributes a small step toward the better application of quantum computers in nonlinear science, with potential for future exploration.

\section{Acknowledgement}
We thank Xianfeng Gu and Nengkun Yu for their useful discussions. This work was supported by the U.S. Department of Energy, Office of Science, Advanced
Scientific Computing Research under Award Number DE-SC-0012704. 
We also acknowledge the support by a Seed Grant from
Stony Brook University’s Office of the Vice President for Research and by the Center for Distributed Quantum Processing.

\bibliography{ref.bib}{}
\bibliographystyle{unsrt}

\clearpage
\newpage
\onecolumngrid

\appendix

In this appendix, we analyze the norm of several quantities that appeared in our algorithms, in order to provide a quantitative guarantee as well as exactness of our algorithm.
\section{Analyzing Norm of Jacobian $|J|$ }
\label{sec: boundJ}
In this section, we show that the Jacobian's norm could be appropriately bounded. Since it is general for arbitrary norm measure, we consider the $l_1$ norm for convenience and simplicity, and we use $|.|$ to refer to such a norm. Since $|J| = |J^T|$, and recall that $l_1$ norm of a matrix is the maximum $l_1$ norm of a column of that matrix, we have that:

\begin{align}
    |J| = \max_{i} | \bigtriangledown f_i (x) |.
\end{align}

We use the following property:
\begin{align}
     | \bigtriangledown f_i (x) | =  | D_i(x) x | \leq |x| |D_i|. 
\end{align}

Assuming that $|x| \leq 1$ (later, we would show that this is also always possible during our iteration procedure), then the norm above is bounded by unity if $|D_i| \leq 1$. According to~\cite{rebentrost2019quantum}, the norm of such gradient operator is bounded as:
\begin{align}
    |D_i| \leq p |A_i|. 
\end{align}
Therefore, as long as $p |A_i| \leq \sqrt{n}$ for all $i$, then the norm of $J$ is guaranteed to be less than $\sqrt{n}$, which means that the norm of $J/\sqrt{n}$ is less than unity. This condition is always possible, for example, one chooses to rescale all matrices $A_i$ in the nonlinear algebraic equations by some factor, e.g., $p$ if the initial matrices have norm less than unity (if they do not, then we can rescale further by its maximum entries). Note that the rescaling trick is very common in many contexts for analysis purposes~\cite{harrow2009quantum, childs2017quantum}, and they do not induce any systematic issues. For example, in our case, adjusting a factor of the set of matrices would not affect the solution. 

\section{Analyzing Norm of $F(x)$}
Recall that:
\begin{align}
    F(x) = \begin{pmatrix}
        f_1 (x) \\
        f_2 (x) \\
        \vdots  \\
        f_n (x) 
    \end{pmatrix}.
\end{align}
Now we take a look in $l_2$ norm:
\begin{align}
    |F(x)|_2 &= \sqrt{  \sum_{i=1}^n |f_i(x)|^2 } \\
    &= \sqrt{  \sum_{i=1}^n | (x^T)^{\otimes p} A_i x ^ {\otimes p}  |^2   }  \\
    & \leq \sqrt{  \sum_{i=1}^n |x|_2^{2p} |A_i|_2   } \\
    & \leq \sqrt{  \sum_{i=1}^n |x|_2^{2p} |A_i|_2^2   } \\
    & \leq \sqrt{  n |x|_2^{2p} |A|_2^2 }  = \sqrt{n} |x|_2^p |A|_2.
\end{align}
As long as $|A|_2 \leq 1 $ we have $|F(x)|_2 \leq |x|_2^p \sqrt{n}$. Throughout the work, we work in the regime $|x| <1$, therefore $|F(x)| < \sqrt{n}$. Again, we remark that the scaling of matrix $A_i$ (for all $i$) is just a matter of convenience, and it would not affect the nature of our nonlinear system, e.g., the solution remains the same.

\section{Proof of Lemma \ref{lemma: 7}}
\label{sec: proof}

To make it easy to follow, we use symbol $A$ and $B$ to denote the Hilbert space corresponds to register $(x x^T)^{\otimes p-1} $ (of dimension $n^{p-1} \times n^{p-1}$ ) and $ \bigtriangledown f_i(x) x^T $ (of dimension $n \times n$) respectively. 

We describe the following procedure: \\ 

Suppose we begin with the state $\ket{\bf 0}_U \ket{\bf 0}_A \ket{\bf 0}_B \ket{k}$ where $\ket{k}$ is a basis in the Hilbert space of dimension $n$. It is easy to use SWAP gates to shuffle the order, i.e., to transform
$$ \ket{\bf 0}_U \ket{\bf 0}_A \ket{\bf 0}_B \ket{k} \rightarrow \ket{\bf 0}_U \ket{k}\ket{\bf 0}_A \ket{\bf 0}_B. $$

Let $\ket{\bf 0}_U$ denotes the corresponding auxiliary qubits required for block encoding the operator $P$ (as we described earlier). We consider the following operation:
\begin{align}
    U \ket{\bf 0}_U \ket{k} \ket{\bf 0}_A \ket{\bf 0}_B  = \ket{\bf 0}_U P \ket{k}\ket{\bf 0}_A  \ket{\bf 0}_B + \ket{\Phi_{\perp}},
\end{align}
where $\ket{\Phi_{\perp}}$ satisfies: $\ket{\bf 0}_U \bra{\bf 0}_U \otimes I \cdot \ket{\Phi_{\perp}} = 0$. 

We take a closer look:
\begin{align}
  P \ket{k}\ket{\bf 0}_A  \ket{\bf 0}_B  = \ket{k} \otimes ( (x x^T)^{\otimes p-1} \ket{\bf 0}_A ) \otimes (  \bigtriangledown f_k(x) x^T \ket{\bf 0}_B  ).
\end{align}
Then, we apply the SWAP gates again between the register $\ket{k}$ and register $A$ to change the order again, i.e., we obtain the following state:
\begin{align}
 ( (x x^T)^{\otimes p-1} \ket{\bf 0}_A ) \otimes \ket{k} \otimes (  \bigtriangledown f_k(x) x^T \ket{\bf 0}_B  ). 
\end{align}

Denote the above \textbf{procedure} as $U_p$. To summarize, $U_p$ acts as following:
\begin{align}
    U_p  \ket{\bf 0}_U \ket{\bf 0}_A \ket{\bf 0}_B \ket{k} &= \ket{\bf 0}_U  (x x^T)^{\otimes p-1} \ket{\bf 0}_A ) \otimes \ket{k} \otimes (  \bigtriangledown f_k(x) x^T \ket{\bf 0}_B + \ket{\Phi_{\perp}}   \\
    &= \ket{\alpha}.
\end{align}

We consider the state: $\ket{\bf 0}_U \ket{\bf 0}_A \ket{\bf 0}_B \ket{i}$ (where $\ket{i}$ is the computational basis). We note that the register $B$ has dimension $n$. Therefore, if we apply the unitary $U_m = I_U \otimes I_A \otimes H^{\otimes \log(n)} \otimes I$ to the state $\ket{\bf 0}_U \ket{\bf 0}_A \ket{\bf 0}_B \ket{i}$, we obtain:
\begin{align}
    \ket{\bf 0}_U \ket{\bf 0}_A (  \frac{1}{\sqrt{n}} \sum_{j=0}^{n-1}\ket{j}  ) \ket{i} = \ket{\beta}.
\end{align}

We remark the following: \\
$$\ket{\bf 0}_U \bra{\bf 0}_U \otimes I \cdot \ket{\Phi_{\perp}} = 0  $$ 
$$\bra{\bf 0}_A (x x^T)^{\otimes p-1} \ket{\bf 0}_A = | \braket{0, x} |^{2(p-1)} = x_0^{2p-2}$$
$$  \frac{1}{\sqrt{n}} \sum_{j=0}^{n-1}\bra{j} \cdot \ket{k} = \frac{1}{\sqrt{n}} $$
$$ \bra{i} \bigtriangledown f_k(x) x^T \ket{0}_B = x_0 ( \bigtriangledown f_k(x) )_i $$
where in the second and last line, we use $x_0 = \braket{0, x}$ to denote the first component of a $n$-dimensional vector $x$. Then we have: 
\begin{align}
    \braket{\beta, \alpha} =  \frac{1}{\sqrt{n}} x_0^{2p-1} ( \bigtriangledown f_k(x) )_i.
\end{align}

According to the definition~\ref{def: blockencode}, the unitary $U_m U_p$ is the block encoding of the matrix:
 ${ x_0^{2p-1} J }/{(\sqrt{n}ps)}$. 
One can reasonably expect that if the value of $x_0$ is very small, then it will incur a substantial running time. A better way to overcome this issue is to change the state $\ket{\bf 0}_A$ in the above construction to some states that are close to any temporal solution $x_k$ during the iteration. As we have mentioned, suppose the initially guessed solution $x_0$ is within the vicinity of the actual solution, then the overlaps between any pairs of $x_0$ and temporal solution $x_k$ (at $k$-th iteration step) are supposed to be sufficiently high, e.g., lower bounded. Therefore, in the above procedure, one simply needs to let unitary $U$ act on the register $A$ to transform the state $\ket{\bf 0}_A$ to the state $\ket{x_0}$ (or $x_0$ as we denoted before) and proceed accordingly. Then the factor $\gamma$ in Lemma \ref{lemma: 7} would be the overlap $|x_0^T x_k|$, which could be estimated using amplitude estimation~\cite{brassard2002quantum, rall2020quantum}. This step only incurs a constant cost overhead and, hence, doesn't increase the overall running time.

\end{document}